
%
%
%
%
\input harvmac.tex

%
%
\lref\haw {S.W. Hawking, Comm. Math. Phys. {\bf 43}  (1975) 199.}
\lref\pres {J. Preskill, Caltech preprint CALT-68-1819 hep-th/9209058}
\lref\wit {E. Witten, Phys. Rev. {\bf {D44}} (1991) 314.}
\lref\hv {H. Verlinde, ``Black Holes and Strings in Two Dimensions'',
 in the proceeding of the Sixth Marcel Grossman Meeting,
 World Scientific (1992).}
\lref\vv {E. Verlinde and H.Verlinde, Nucl. Phys. {\bf B 406} (1993) 43}
\lref\svv {K. Schoutens, E. Verlinde and H. Verlinde, Phys. Rev. {\bf D 48}
(1993) 2690.}
\lref\cghs {C. Callan, S. Giddings, J. Harvey, and A. Strominger,
 Phys. Rev. {\bf{D45}} (1992) 1005.}
\lref\rst {J.G. Russo, L. Susskind, L. Thorlacius, Phys. Rev. {\bf D47}
(1993) 533}
\lref\jhs {J. Harvey and A. Strominger, in {\it String Theory and
 Quantum Gravity}, Proceedings of 1992 Trieste Spring School,
 (World Scientific, 1993).}
\lref\str {A. Strominger, Phys. Rev. {\bf D 46} (1992) 4396.}
\lref\cj {D. Cangemi and R. Jackiw, MIT preprint, CTP 2165, hep-th/9302026
(February 1993)}
\lref\thooft{G. 't Hooft, Nucl. Phys. {\bf B335} (1990) 138.}
\lref\msw {G. Mandal, A Sengupta, and S. Wadia, Mod. Phys. Lett. {\bf A6}
(1991) 1685}
\lref\withole {E. Witten, Phys. Rev. {\bf{D44}} (1991) 314}
\lref\bical{A. Bilal and C. Callan, Nucl. Phys. {\bf B 394} (1993) 73.}
\lref\DDF{E. Del Giudice, P. Di Vecchia and S. Fubini, Ann. Phys. {\bf 70}
(1972) 378.}
\lref\comp{L. Susskind, L. Thorlacius and J. Uglum, Stanford preprint
SU-ITP-93-15.}
\lref\brower{R.C. Brower, Phys. Rev. {\bf D6} (1972) 1655.}
\lref\gn {S. Giddings and W. Nelson, Phys. Rev. {\bf D 46} (1992) 2486.}
\lref\hks {S. Hirano, Y. Kazama, Y. Satoh, preprint UT-Komaba 93-3}
\lref\birdav {See e.g. N. Birrell and P. Davies, {\it Quantum Fields in
Curved Space} (Cambridge, 1982) and references therein.
For more recent discussions of the moving mirror-black hole analogy see
R. Carlitz and S. Willey, Phys. Rev {\bf D 36} (1987) 2327, 2336, and
F. Wilczek, IAS-preprint HEP-93/12, hep-th/9302096 (February 1993).}
\lref\cv {T. Chung and H. Verlinde, Nucl. Phys. {\bf B418} (1994) 305,
 hep-th/9311007.}
\lref\cthesis {T. Chung, Ph.D. thesis, Princeton University (1994).}
\lref\kvv {Y. Kiem, E. Verlinde, H. Verlinde, private communication.}


\def\X{{\rm \, x}}


\Title{\vbox{\baselineskip12pt\hbox{NHCU-HEP-94-29}
    \hbox{October 1994}
    \hbox{hep-th/9410155}
    }}
{\vbox{\centerline{Unitarity, Analytical Time}
	\vskip 2mm\centerline{and Hawking's Temperature}}}

\vskip .35cm
\centerline{Tze-Dan Chung\footnote{$^\dagger$}
{\it email: chung@cc.nctu.edu.tw} }
\bigskip\centerline{\it Department of Electrophysics}
\centerline{\it National Chiao Tung University}
\centerline{\it Hsinchu, Taiwan 30050}

\vskip 1.2cm
\rm
\noindent

\centerline{\bf Abstract}

By analytically continuing the time variable in a black hole
background, and requiring unitary evolution, it is found that
quantum mechanical states at the horizon develop a thermal factor
under suitable identification of the physical time.
The thermal factor is found
to be determined exactly by the Hawking's temperature.
This can be interpreted as the Hawking's radiation and offers an
alternative understanding to the process and the information loss
puzzle.

\Date {}


\newsec {Introduction}

It has been about twenty years since Hawking \refs{\haw}
first put forward his argument
of black hole radiation. In his semi-classical calculation, it was shown that
pair creation in black holes produces thermal radiation. The temperature of
radiation is inversely proportional to the mass of the black hole.
Therefore, a black hole will radiate at an ever-increasing rate until
it finally reaches the Planck mass where the semiclassical approximation
breaks down.
This implies information loss and destruction of quantum coherence.
Physicists have been puzzled by this phenomenon since then.
Efforts in the past has not been successful in resolving this paradox.
Perhaps we need to look at the problem from a completely new perspective.

The relation between thermodynamics and quantum mechanics was realized soon
after quantum field theory was developed. It was found that the partition
functions in thermodynamics and those in quantum field theories are related
by a phase rotation of $\pi \over 2$ in the complex plane of the time
variable (or $1\over k_B T$). Hence, there is a corresponding result in
thermodynamics for each calculation in quantum field theory, and vice versa.
Moreover, a thermal background can be created by shifting the time by
an imaginary number.
We also know that there is no statistical
mechanics in purely classical physics, since counting the number of states
needs the energy levels of a system to be quantized.
It is, therefore, not unnatural to speculate
that these two seemingly different
branches of physics are somehow related to each other.

With the aforementioned borne in mind,
in this paper we will study the analytical continuation of the time
variable in a black hole background and look for its relation to the
thermodynamical properties of the black hole.
It is found that, by requiring unitary evolution,
a quantum mechanical state escaping from the horizon
will change its norm when described by the physical time.
The new norm of the state differs from the original
one by a thermal factor of Hawking's temperature.
This is interpreted as the Hawking's radiation process.

We will first present the argument in the two dimensional black hole case.
Then we will repeat the same procedure to the four dimensional black holes.
The arguments are on parallel lines.
The dilaton black hole model serves as an illustration and warm up exercise,
while considering the Kruskal coordinates in the four dimension case justifies
our definition of the physical time.
However, the calculation in the four dimensional
case is found to be just as simple, although
most of the motivations are initiated from studying the two dimensional model.

\newsec{Energy and time in the complex plane}

In traditional quantum mechanics, we define the Hamiltonian operator as
complex valued, in general. We also restrict the time to be real.
We then examine the condition for the theory to be physical,
either by requiring energy conservation or unitary time evolution.
Unitarity implies that the time translation operator
\eqn\tiev{U=e^{iHt}}
is unitary, i.e., $UU^{\dagger}=1$.
This in turn requires the Hamiltonian $H$ to be
hermitian. What we are trying to accomplish in this paper
is to relax the restriction
on the time $t$, by also allowing it to be complex valued.
The motivation, and justification, for doing this will be explained soon.
At the same time, we still require unitary ``time'' evolution.
Now $H$ alone may not be hermitian, instead, the product $Ht$ is required to
be hermitian if our theory is to be {\it physical}.

Now we turn to the discussion on the time $t$.
We know that almost all the results
of quantum field theory can be translated into that of thermodynamics,
simply by analytically continuing the ``time'' into the complex plane
(in fact, the imaginary axes). Therefore, it is not surprising at
all if we start to talk about complex ``time'' by analytical continuation
and find some thermal properties at the end.
In general relativity, this becomes most transparent when we consider the
black hole solution of the Einstein's equation in the Kruskal coordinates,
when we continue the Kruskal time {\it into} the horizon ($r<2M$).

The crucial point of our argument lies in the identification
of the physical time. We propose that the fiducial observer always sees the
time as a real number.
As one looks beyond the horizon, it is inevitable to make measurements
with physical time, as otherwise there is no communication with the
asymptotic observer.
However, the Hamiltonian stays the same and Hawking's radiation follows
as a result of non-hermiticity.
This will be made more precise as we study the examples in the following
sections. A different identification of the physical time certainly
produces different physics. We will also discuss about this possibility
later on.

\newsec{Hawking's radiation in dilaton black holes}

Now let us see what happens when we consider fluctuations at the boundary
of the two dimensional dilaton black hole.
We will work in the model and use the convention as in \refs{\cv},
where the boundary of the black hole space time is described by the
trajectory $\X^+ (\X^-)$. The variables $\X^+,\X^-$ are called Kruskal
coordinates. They are related to the fiducial (Minkowski) coordinates by
\eqn\krus{\lambda \X^{\pm} = \pm e^{\lambda (r \pm t)}, }
with $\lambda$ being the cosmological constant.
We see here that for real ($r,t$), $\X^{\pm}$ resides in the fourth quadrant
$(+,-)$. A typical black hole solution has boundary trajectory asymptote
of the form
\eqn\traj{ -\lambda^2 \X^+ \X^- = A}
near the future infinity $\cal I ^+$,
where $A$ is a positive constant. The horizon is the line $\X^- =0$.
Studies show that the trajectory turns space-like at certain incoming
energy flux \refs{\rst} \refs{\cv},
where the horizon in Kruskal coordinate $\X^-$ is shifted correspondingly.
Black hole sets in when $\X^- > 0$ and information is lost.
It was unclear how to deal with waves propagating beyond the horizon, and
it was concluded that information is loss.
But suppose we do ``look'' beyond the horizon, what will we see?


It was found \refs{\svv} \refs{\cv} that there is a horizon shift
induced for incoming matter flux. Therefore, considering a small shift in
the horizon is equivalent to considering perturbation from the incoming
matter flux. Because the position of the horizon is dynamically
determined by the amount of incoming energy flux.

Let us suppose that there is a shift $\Delta$, where $\Delta>0$,
i.e., before making the shift,
the coordinate $\lambda \X^-$ runs from $- \infty$ to $\Delta$,
instead of $- \infty$ to $0$.
In two dimensional models this shift shows that a black hole has been formed.
The shift is determined by requiring that the future infinity ${\cal I} ^+$
corresponds to the point $(\X^+, \X^-)=(\infty,0)$
\footnote{$^*$}{In the model of \refs{\cv}, it
corresponds to the super-critical case where the incoming massless scalar
field energy is large enough to make a temporal black hole.}.
Then some of the matter that formed the black hole must have travelled
to a space-time point with a positive value of $\lambda \X^-=\delta < \Delta$.
The magnitude of the value $\delta$ indicates how far beyond the horizon it
has traversed. In this paper it suffices to consider small $\delta$.

Look at the space-time point $\lambda \X^- = \delta$, with
$0< \delta < \Delta$. Considering fluctuations just beyond the horizon means
we are looking at small $\delta$. Equivalently, we can consider a virtual
state created right beyond the horizon ($r=0$), and observe its propagation
(aka, tunneling) through it.
In the fiducial coordinates this is the point
\eqn\fidst{ t = - {1\over \lambda} [log {\delta }+ \pi i].}
Requiring unitary time evolution, we have
\eqn\unita{Ht=H^{\dagger} t^*}
Let
\eqn\hamri{H=E+\epsilon i ,}
unitarity condition \unita \ implies that
\eqn\eps{\epsilon = -{ \pi E \over log {\delta }}}.

Next we will identify the physical time.
Since the physical time is an observable, it should be real valued.
Define the physical time $t_{phy}$ as
\eqn\phykru{t_{phy}= - {1\over 2 \lambda} log {(|\X^+ / \X^-|) }.}
This reduces with \krus \ for negative $\X^-$.

Noting that $\lambda \X^+= \delta ^{-1}$ at the horizon $r=0$,
we then have the physical time beyond the horizon as
\eqn\phyt{ T = - {1\over \lambda} log {\delta }.}
We see here that $\delta =0$ corresponds to $T \to - \infty$,
so that $T$ increases with $\delta$.
Note also that $\delta =0$ is also the point
$\X^- = 0$, or $t \to \infty$. Therefore, the direction of time flow
is preserved. $T$ starts at where $t$ ends.

Let us now consider a quantum mechanical state from behind the horizon.
We will use the analytical Hamiltonian $H$, and the physical time $T$.
One way of perceiving this is by comparing it to the
``uncertainty principle''.  To construct the time evolution operator we
need to know both $H$ and $t$. Once $t$ is determined (to be the physical
time $T$), $H$ is ``uncertain'' up to a phase.
It is equivalent to determine $H$ first and we will arrive at the same result.
But for any experiment to be done it is better to have a well-defined time.

The evolution of the state
$|\psi>$ is as follows:
\eqn\stev{ | \psi(T)>= e^{-iHT} | \psi(-\infty)>.}
Putting in $H$ and $T$ we find
\eqn\psit{|\psi(T)>= e^{-iEt}e^{ -{\pi E T \over log \delta}}|\psi(-\infty)>.}
This means that
\eqn\normt{ <\psi(-\infty)| \psi(-\infty)> = e^ { -{2 \pi E \over \lambda}}
<\psi(T)|\psi(T)>}
This result shows that any state,
when passing through the boundary, has its norm decreased by a thermal factor
of the Hawking's temperature $k_B T_H = {\lambda \over 2 \pi}$.
In other words, pair creation from the horizon is thermal.
There is something very special about \normt \ : it does not depend
on the ``time'' $T$. This implies that the same effect occurs for tunneling
from different sites of $T$. In classical mechanics we do not have wave
travelling backward in time as in this case (from $T$ to $-\infty$),
but it is perfectly legitimate in quantum mechanics, when tunneling effect
occurs.

How should we interpret this in terms of particle creation?
The particle (virtual, from the perspective of the fiducial observer)
has made its way (tunneled) through the horizon.
We do not expect to see this normally,
since it would take an infinite amount of time for anything to achieve this.
Because the time $T$ describes the space-time beyond
the horizon, which is causally disconnected from any asymptotic observer.
However, the fact that the amplitude does not depend on $\delta$ indicates
that it may just have reappeared at the point $\delta=0$, overcoming the
infinite time delay.
This means that, in the future infinity
$\cal I ^+$ ($\X^- = \delta =0$), thermal radiation of temperature $T_H$
is detected.
Here we need to clarify one more point. For propagation on the same side
of the horizon the norm of the state is $1$, since it is possible to define
a real time and a hermitian Hamiltonian.
But that does not tell us anything because there are causally disconnected
space-time, and irrelavant to our discussion here.
There is a phase change when a black hole is formed.
The fact that our result is $T$ independent means that the creation is not
a local one. Instead, particles created right beyond the horizon
immediately appear on the horizon, where $\delta=0$, or $T=-\infty$.
And all of them are thermal, with temperature $T_H$.

What has happened to the particle? There is one plausible explanation.
For any particle trying to make its way out of the horizon, a
fraction of it ($1- {1\over k_B T_H}$) stays in the black hole,
only a fraction $1\over k_B T_H$ managed to appear on the other side of
the horizon.
There is no violation to quantum mechanics since unitarity is {\it a priori}.
But there is Hawking's radiation
in the horizon of a black hole. For the ``super''-observer, the scattering
matrix is still unitary, since we have required $U$ to be unitary, in terms
of his analytical time $t$ and $H$.

As a last comment,
many other ways of identifying the physical time produce
the same result, although we believe ours is the correct one.
 For example, the same physical time can be derived if we define $T$ to be the
real part of the complex time $t$.

\newsec{Four dimensional black holes}

Now we discuss the realization of the results in the four dimension.
The assumptions and arguments in this case is parallel
to the two dimensional one. Therefore, here we will keep it short and
concise, by avoiding redundancy in argument.

The Kruskal coordinates in four dimensional black holes is defined as:
\eqn\krfd{\eqalign{ \X^+ \X^- &= -16 M^2( {r \over 2M} -1 ) e^{r/2M}  \cr
 \X^+ / \X^- &= - e^{t /2M }  },}
in which the Schwarzchild's metric is
\eqn\schwa{ds^2 = -{2M \over r} e^{-{r\over 2M}} d\X^+ d\X^- + r^2 d\Omega^2}
Here we see that real values of
$(\X^+, \X^-)$ resides in the fourth quadrant where $r > 2M$.
For $r<2M$, i.e., beyond the horizon, we have to analytically continue
the Kruskal coordinates into the complex plane.
The same prescription for the physical time beyond the horizon as the previous
section divides $t$ into two coordinate patches. The first being the usual
definition \krfd \ where $t$ runs from $-\infty$ to $\infty$. This is glued
to the second coordinate patch with $| \X^+ / \X^-| =  e^{T /2M}$ at the
horizon. From there time $T$ is from $-\infty$ to $\infty$.
$T$ starts at where $t$ ends.
Therefore, the physical time defined in the previous section is, in fact,
``physical''.

Studies have also shown that there are similar horizon shifts
in the Kruskal coordinates as in the two dimensional case when a black
hole has been formed \refs{\kvv}.
This further justifies our assumption that horizon is not a sharply defined
line and we can look beyond it by analytical continuation.

Therefore, we can immediately write down the Hawking's temperature for the
four dimensional black holes. By comparing the definitions of the Kruskal
coordinates we see that we have to replace $\lambda$ with $ 1\over 4M$,
where $M$ is the mass of the black hole.
The Hawking's temperature for the four dimensional black holes is thus
found to be $k_B T_H = {1 \over 8 \pi M}$, exactly the same as
what was already known.

\newsec{Discussion}

To summarize, there are two assumptions in our calculation, and one
interpretation.

1. The product $Ht$ is hermitian for unitarity requirement.

2. When we look beyond the horizon, once the the physical time $T$ is
identified, the analytical Hamiltonian $H$ is used.

3. Any virtual state created beyond the horizon, if it is to escape
from the black hole and emerge to the observer's side, acquires a thermal
factor in its norm. This is interpreted as the Hawking's radiation.

It is not fully understood yet how assumptions 1 and 2 apply to other
branches of physics. But they are very plausible assertions
since we have already seen the striking similarity
between statistical mechanics and quantum field theories.

If our interpretation 3 is valid, Hawking's radiation is simply an effect of
coordinate choice. To make any measurement we are forced to
use one coordinate for the physical time,
and another for the Hamiltonian.
It is very similar to 't Hooft's \refs{\thooft}
argument involving the infinite red-shift between the local and
asymptotic time.
Hawking's radiation, in this picture, is a {\it result} of requiring
unitary time evolution, instead of violating it.
In the correct prescription of energy and time, unitarity is always
preserved and we will never see Hawking's radiation. But the time for such
observer is not a physical observable.
If, instead, we try to measure physics in the ``physical'' time $T$,
we will detect thermal radiation
as predicted by Hawking.
The reason for our seeing the infinite amount of radiation from black holes
is merely an artifact of coordinate choice.


It could also be argued that dimensional analysis may lead to similar
results. However, our results are exact, while dimensional argument
is always up to a constant. Moreover, we have provided a possible physical
origin for these effects.

At last, we should make it clear that this is only a tentative calculation.
The physical origin of the assumptions and the validity of our interpretation
of the result requires further investigation.
We are convinced that it worths
being written down because of its simplicity, and the surprising coincidence
of the result with Hawking's calculation.
Even if this were not true, the calculation itself offers another aspect
of studying the thermodynamical property of black holes,
and a new way of looking at the problem.

\vfill\eject

\noindent
{\bf Acknowledgements.}

I would like to thank H. Verlinde for initiating my research in the
black hole radiation puzzle.
I would also like to thank the Institute of Applied Mathematics of NCTU for
allowing me access to its computer facilities.
This research was financially supported by National Science Council of Taiwan,
R.O.C., under grant number NSC84-2811-M009-006.

\listrefs

\end